# Speed and Energy Optimised Quasi-Delay-Insensitive Block Carry Lookahead Adder

P. Balasubramanian [§, *], D.L. Maskell [§], and N.E. Mastorakis [ξ]

[§] School of Computer Science and Engineering

Nanyang Technological University

Singapore 639798

[ξ] Department of Industrial Engineering

Technical University of Sofia

Sofia 1000, Bulgaria

* Corresponding author. E-mail: balasubramanian@ntu.edu.sg

## Abstract

We present a new asynchronous quasi-delay-insensitive (QDI) block carry lookahead adder with redundancy carry (BCLARC) realized using delay-insensitive dual-rail data encoding and 4-phase return-to-zero (RTZ) and 4-phase return-to-one (RTO) handshaking. The proposed QDI BCLARC is found to be faster and energy-efficient than the existing asynchronous adders which are QDI and non-QDI (i.e., relative-timed). Compared to existing asynchronous adders corresponding to various architectures such as ripple carry adder (RCA), conventional carry lookahead adder (CCLA), carry select adder (CSLA), BCLARC, and hybrid BCLARC-RCA, the proposed BCLARC is found to be faster and more energy-optimised. The cycle time (CT), which is the sum of forward and reverse latencies, governs the speed; and the product of average power dissipation and cycle time viz. the power-cycle time product (PCTP) defines the low power/energy efficiency. For a 32-bit addition, the proposed QDI BCLARC achieves the following average reductions in design metrics over its counterparts when considering RTZ and RTO handshaking: i) 20.5% and 19.6% reductions in CT and PCTP respectively compared to an optimum QDI early output RCA, ii) 16.5% and 15.8% reductions in CT and PCTP respectively compared to an optimum relative-timed RCA, iii) 32.9% and 35.9% reductions in CT and PCTP respectively compared to an optimum uniform input-partitioned QDI early output CSLA, iv) 47.5% and 47.2% reductions in CT and PCTP respectively compared to an optimum QDI early output CCLA, v) 14.2% and 27.3% reductions in CT and PCTP respectively compared to an optimum QDI early output BCLARC, and vi) 12.2% and 11.6% reductions in CT and PCTP respectively compared to an optimum QDI early output hybrid BCLARC-RCA. The adders were implemented using a 32/28nm CMOS technology.

# 1. Introduction

The 2017 edition of the International Roadmap for Devices and Systems [1] suggests that asynchronous design could be a potential solution to address the increasing power/energy consumption of a digital circuit/system. Substantiating this, in [2], a 128-point, 16-bit, radix-8 fast Fourier transform (FFT) processor was implemented in the robust QDI asynchronous design style and it was compared with a conventional synchronous FFT processor implementation, and both were realized using a 65nm CMOS technology. It was noted that, overall, the QDI FFT processor is 34× more energy-efficient than its synchronous equivalent. The QDI design style is a promising alternative to the synchronous design style, and different types of QDI implementations exist.

QDI circuits are known to be robust to process, voltage, timing and temperature variations [3, 4], which is important to note since the issue of variability [5] is quite common in the nanoelectronics era. Moreover, QDI circuits are less affected by electromagnetic interference compared to synchronous circuits [6]. These properties make QDI circuits preferable for secure applications [7, 8]. Further, QDI circuits and systems are modular [9], and hence they are convenient to reuse/replace thus obviating the need for extensive timing re-runs and analysis. Furthermore, QDI circuits are naturally elastic [10] unlike synchronous circuits, and they are suitable for the subthreshold operation [11].

A QDI circuit is the practically realizable delay-insensitive circuit which includes the weakest compromise of the isochronic fork [12]. The isochronic fork assumption implies that all the wires branching out from a node/junction would experience concurrent rising or falling signal transitions. Usually, the isochronic fork assumption is confined to a small circuit area and hence their realization would not be difficulty. It has been shown in [13] that QDI circuits are realizable in the nano-electronics regime.

Addition is a fundamental operation in computer arithmetic, which is realized using the adder, and an effective adder design is of interest and importance. This article deals with the high-speed and energy-efficient QDI realization of the adder. In a latest work [14], several asynchronous implementations of a 32-bit adder were considered and analysed. In this work, we propose a QDI BCLARC that outperforms all the asynchronous adders discussed in [14] and [15] with respect to speed (CT) and energy (PCTP).

The rest of the article is organized as follows. Section 2 gives the nomenclature. Section 3 discusses the design preliminaries of QDI and non-QDI asynchronous circuits. By non-QDI circuits, we refer to relative-timed circuits [16]. Section 4 describes the proposed QDI sub-BCLA block without and with the redundant carry output and the resulting QDI BCLARCs by considering an example 32-bit addition. Section 5 presents the design metrics estimated for several 32-bit asynchronous adders corresponding to 4-phase RTZ and 4-phase RTO handshaking, and they are compared. Finally, Section 6 draws some conclusions.

## 2. Nomenclature

Widely used acronyms and their expansions are given below for a ready reference.

- CLA – Carry Lookahead Adder

- BCLA – Block CLA

- BCLARC – BCLA with Redundant Carry

- BCLG – Block Carry Lookahead Generator

- BCLGRC – BCLG with Redundant Carry

- CCLA – Conventional CLA

- CSLA – Carry Select Adder

- CT – Cycle Time

- PCTP – Power-Cycle Time Product

- RCA – Ripple Carry Adder

- QDI – Quasi-delay-insensitive

- RTO – Return-To-One

- RTZ – Return-To-Zero

# 3. QDI and Non-QDI Circuits – A Background

The design fundamentals of QDI and non-QDI (relative-timed) asynchronous circuits are briefly discussed in this section to provide a background.

## 3.1. Data Encoding, Handshaking, and Timing Parameters

The general schematic of a QDI or a relative-timed circuit stage encompassing delay-insensitive data encoding and a 4-phase handshaking is shown in Fig. 1a based on the transmitter-receiver analogy. The corresponding technical schematic is shown in Fig. 1b.

In Fig. 1b, the current stage and next stage registers are analogous to the transmitter and the receiver shown in Fig. 1a, and a QDI or a relative-timed circuit is sandwiched between the current stage and the next stage register banks. The register bank comprises a series of registers, with one register allotted for each of the rails of a dual-rail encoded data input. The register refers to a 2-input Muller C-element [17]. The C-element will output 1 or 0 if all its inputs are 1 or 0 respectively. If the inputs to a C-element are not identical then the C-element would retain its existing steady-state. The circles with the marking 'C' represent the C-elements in the figures.

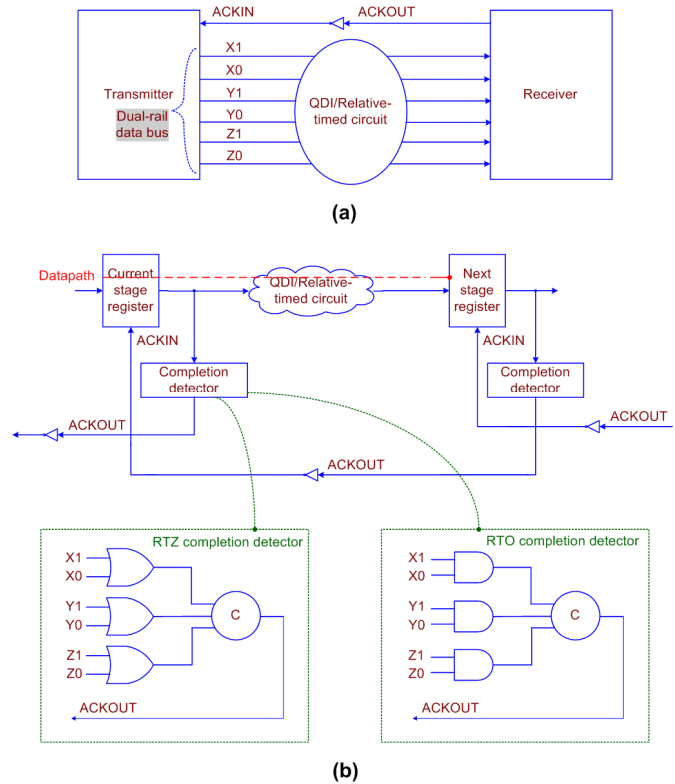

**Fig 1. (a) Transmitter-Receiver analogy of a QDI/non-QDI (relative-timed) asynchronous circuit stage, and (b) technical schematic portraying the example RTZ and RTO completion detectors for the presumed dual-rail data bus comprising inputs (X1, X0), (Y1, Y0) and (Z1, Z0). The OR and AND gates used in RTZ and RTO completion detectors are the duals of each other. The datapath is highlighted by the red dashed line in (b).**

In Fig. 1, (X1, X0), (Y1, Y0) and (Z1, Z0) represent the dual-rail encoded primary inputs of the corresponding single-rail inputs X, Y and Z. According to delay-insensitive dual-rail data encoding and the 4-phase RTZ handshaking [9], an input W is encoded as (W1, W0) where W = 1 is represented by W1 = 1 and W0 = 0, and W = 0 is represented by W0 = 1 and W1 = 0. Both these assignments are called *data*. The assignment W1 = W0 = 0 is called the *spacer*, and the assignment W1 = W0 = 1 is deemed illegal since the coding scheme should be complete [18] and unordered [19] to maintain the delay-insensitivity.

The application of input data to a QDI or a relative-timed circuit which adheres to the 4-phase RTZ handshaking follows the sequence of *data-spacer-data-spacer*, and so forth. It may be noted that the application of data is followed by the application of the spacer, which implies that there is an interim RTZ phase between the successive applications of input data. The interim RTZ phase ensures a proper and robust data communication i.e., handshaking between the transmitter and the receiver. The RTZ handshake protocol is specified by the following four steps:

- First, the dual-rail data bus specified by (X1, X0), (Y1, Y0) and (Z1, Z0) assumes the spacer, and therefore the acknowledgment input (ACKIN) is equal to binary 1. After the transmitter transmits a data, this would cause rising signal transitions i.e., binary 0 to 1 to occur on one of the dual rails of the dual-rail data bus

- Second, the receiver would receive the data sent and drive the acknowledgment output (ACKOUT) to 1. ACKIN is the Boolean complement of ACKOUT and vice-versa

- Third, the transmitter waits for ACKIN to become 0 and would subsequently reset the dual-rail data bus, i.e., the dual-rail data bus assumes the spacer again

- Fourth, after an unbounded (but a finite and positive) time duration, the receiver would drive ACKOUT to 0 and then ACKIN would assume 1. With this, a single data transaction is said to be completed and the asynchronous circuit is permitted to start the next data transaction

According to dual-rail data encoding and the 4-phase RTO handshaking [20], an input V is encoded as (V1, V0) and V = 1 is represented by V1 = 0 and V0 = 1, and V = 0 is represented by V0 = 0 and V1 = 1. Both these assignments are called *data*. The assignment V1

= V0 = 1 is called the *spacer*, and the assignment V1 = V0 = 0 is deemed illegal to maintain the delay-insensitivity.

The application of input data to a QDI or a relative-timed circuit conforming to the 4-phase RTO handshaking follows the sequence of *spacer-data-spacer-data*, and so forth. It may be noted that there is an interim RTO phase between the successive applications of input data. The interim RTO phase ensures a proper and robust data communication between the transmitter and the receiver. The RTO handshaking process is specified by the following four steps:

- First, ACKIN is equal to binary 1. After the transmitter transmits the spacer, this would cause rising signal transitions i.e., binary 0 to 1 to occur on all the rails of the dual-rail data bus

- Second, the receiver would receive the spacer sent and drive ACKOUT to 1

- Third, the transmitter waits for ACKIN to become 0 and would then transmit the data through the dual-rail data bus

- Fourth, after an unbounded (but a finite and positive) time duration, the receiver would drive ACKOUT to 0 and subsequently ACKIN would assume 1. With this, a single data transaction is said to be completed and the asynchronous circuit is permitted to start the next data transaction

In a QDI or a relative-timed circuit, the time taken to process the data in the datapath, highlighted by the red dashed line in Fig. 1b, is called *forward latency*, and the time taken to process the spacer is called *reverse latency*. Since there is an intermediate RTZ or RTO phase between the application of two input data sequences, the *cycle time* is given by the sum of forward and reverse latencies. The cycle time of a QDI or a relative-timed asynchronous circuit is the equivalent of the clock period of a synchronous circuit. The cycle time governs the speed at which new data can be input to an asynchronous circuit.

The gate-level details of example completion detectors corresponding to RTZ and RTO handshaking is shown at the bottom of Fig. 1b, within the dotted green boxes. The completion detector indicates i.e., acknowledges the receipt of all the primary inputs given to an asynchronous circuit stage. In the case of 4-phase RTZ handshaking, ACKOUT is produced using a 2-input OR gate to combine the respective dual rails of each encoded primary input and synchronizing the outputs of all the 2-input OR gates using a C-element or a tree of C-elements. In the case of 4-phase RTO handshaking, ACKOUT is produced using a 2-input AND gate to combine the respective dual rails of each encoded primary input and subsequently synchronizing the outputs of all the 2-input AND gates using a C-element or a tree of C-elements.

## 3.2. QDI Circuits

QDI circuits are classified into three types as strong-indication [21], weak-indication [21], and early output [22] circuits. The input-output timing relations of QDI circuits are illustrated by the representative timing diagrams in Figs. 2a and 2b with respect to RTZ and RTO handshaking.

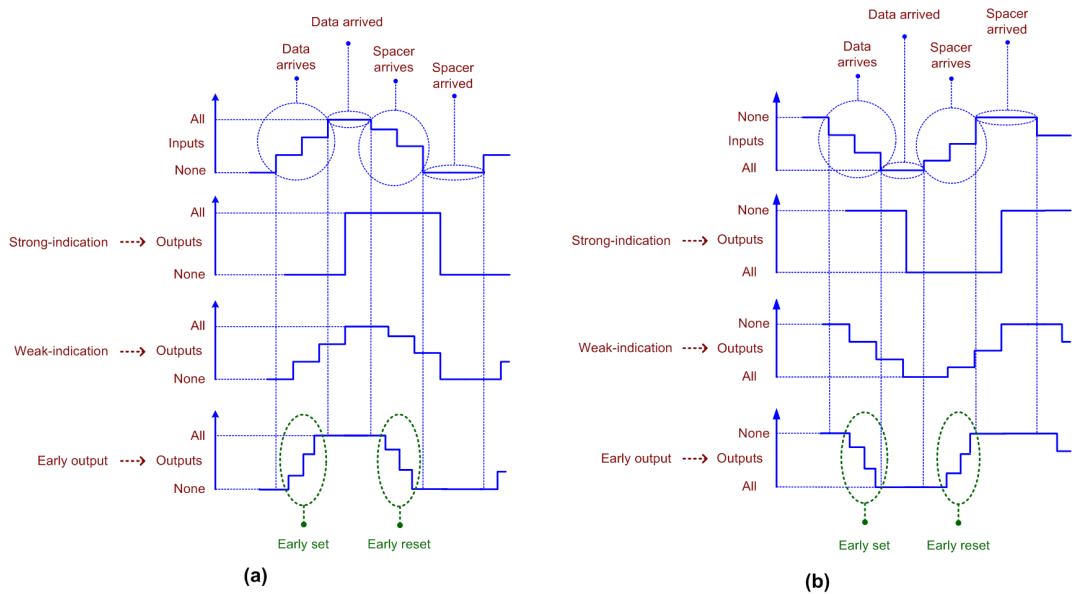

**Fig 2. Input-output timing relation of different types of QDI circuits corresponding to (a) RTZ handshaking, and (b) RTO handshaking. The early set and reset behaviours of the early output circuit type are highlighted by the dotted green ovals in (a) and (b).**

Strong-indication circuits would wait to receive all the primary inputs (data and spacer), and after receiving them would process them to produce the required primary outputs (data and spacer respectively). On the other hand, weak-indication circuits can produce all but one of the primary outputs after receiving a subset of the primary inputs. Nevertheless, only after receiving the last primary input, they would produce the last primary output.

A connection of strong-indication sub-circuits may not result in a strong-indication circuit; rather, a weak-indication circuit may result. For example, if two strong-indication full adders are connected, it could result in a weak-indication 2-bit RCA. This is because if all the inputs to one of the full adders are provided, the corresponding sum and carry output bits of that full adder could be produced regardless of the non-arrival of the inputs to the other full adder in the RCA. However, only after all the inputs to the other full adder are provided, its corresponding sum and carry output bits would be produced. This scenario is characteristic of weak-indication.

While comparing strong- and weak-indication circuit types, the latter are preferable [23], and this is because of the strict timing restrictions inherent in the former. Especially, for implementing arithmetic functions, the weak-indication type is preferable to the strong-indication type and this is due to the following reasons: i) strong-indication arithmetic circuits tend to encounter worst-case forward and reverse latencies for the application of data and spacer, and therefore the cycle time of strong-indication arithmetic circuits is always the maximum (worst-case timing), ii) weak-indication arithmetic circuits may encounter data-dependent forward and reverse latencies or a data-dependent forward latency and a constant reverse latency, and so the cycle times of weak-indication arithmetic circuits are usually less compared to those of strong-indication arithmetic circuits.

An early output circuit is however more relaxed compared to strong- and weak-indication counterparts. After receiving a subset of the primary inputs (data or spacer), an early output circuit can produce all the primary outputs (data or spacer respectively). This implies the late arriving primary inputs may not be acknowledged by the circuit. However, this does not cause any concern because isochronic fork assumptions are imposed on all the primary inputs, and all the primary inputs are provided to the completion detector that precedes the early output circuit, as seen in Fig. 1b. Hence, the acknowledgment of the late arriving primary inputs by the completion detector also implies the receipt of those primary inputs by the asynchronous circuit. Thus, the problem of wire orphan(s) i.e., unacknowledged signal transitions on the wire(s) due to the late arriving input(s) is overcome through the assumption of isochronic forks, which is imposed on all the primary inputs.

Either the data may be produced early, or the spacer may be produced early in an early output circuit. Accordingly, an early output circuit is categorized as early set or early reset kind. The early set and early reset behaviours of early output circuits are highlighted by the dotted green ovals in Figs. 2a and 2b. An early output RCA is preferable to a strong- and a weak-indication RCA for achieving optimizations in speed and power/energy [24]. In general, an early output circuit can achieve enhanced optimizations in the design metrics compared to strong- and weak-indication counterparts.

In a QDI circuit, the logic decomposition should be performed safely [25]. Safe QDI logic decomposition [26] is essential to avoid the problem of gate orphans, which are unacknowledged signal transitions occurring on the intermediate gate output(s). For an illustration of gate and wire orphans, the interested reader is referred to [27]. However, we discuss about orphans in the following section.

The signal transitions will have to occur monotonically throughout an entire circuit from the first logic level, which receives the primary inputs, up to the last logic level, which produces the primary outputs in a QDI circuit [28]. The signal transitions should be either rising or falling throughout an entire QDI circuit. In general, the signal transitions will be rising (i.e., binary 0 to 1) for the application of data and falling (i.e., binary 1 to 0) for the application of spacer in a QDI circuit that corresponds to RTZ handshaking. On the other hand, the signal transitions will be rising for the application of spacer and falling for the application of data in a QDI circuit that corresponds to RTO handshaking.

For monotonicity of signal transitions, the monotonic cover constraint [9] should be incorporated into a QDI logic description. For example, this implies if a QDI logic function is expressed in the sum-of-products form, only one product term should be activated for the application of an input data, i.e., the product terms comprising the sum-of-products expression of a QDI logic function should be mutually orthogonal (also called disjoint), and the logical conjunction of any two product terms in a QDI logic function should yield zero. Thus, a QDI logic function is ideally expressed in the disjoint sum-of-products form [29], which would consist of mutually disjoint product terms to satisfy the monotonic cover constraint. An example illustration of the monotonic cover constraint is given in Section 2.2 of [14], and an interested reader may refer to the same. Embedding the monotonic cover constraint and performing safe QDI logic decomposition are central to the correct implementation of a QDI circuit.

Incorporating the monotonic cover constraint in a QDI logic function would ensure the activation of just one signal propagation path from a primary input to a primary output for the application of input data. This is useful to facilitate the proper acknowledgment of signal transitions throughout an entire QDI circuit, thus avoiding the likelihood of any gate orphan occurrence(s). Gate orphans are troublesome unlike wire orphans as they may affect the robustness of a QDI circuit and if they are imminent, restricting them from affecting the circuit robustness may require incorporating additional timing assumptions which are likely to be sophisticated and hence may be practically difficult to realize [22].

### 3.3. Relative-Timed (Non-QDI) Circuits

Relative-timed circuits [16] are not QDI circuits although they may embed the monotonic cover constraint and adopt safe QDI logic decomposition for their physical realization. This is because relative-timed circuits tend to incorporate extra timing assumptions (in addition to the assumption of isochronic forks), to eliminate any potential problem due to the gate orphan(s). Usually, the extra timing assumptions are related to the delayed arrival of some internal input signals, subject to a specified time bound. If the extra timing assumptions are upheld in a relative-timed circuit they would appear to be QDI, and supposing they are violated, they would not be QDI. Relative-timed circuits are early output circuits; however, they are non-QDI unlike the latter. A couple of relative-timed RCAs were presented in [15], which were realized using early output full adders. Relative-timed circuits are seen to be competitive to early output QDI circuits as they could pave the way for enhanced optimizations of the design metrics compared to strong-indication, weak-indication and early output QDI circuits but at the expense of some compromise in the robustness. Hence, only strong-indication, weak-indication and early output QDI circuits are robust and have freedom from the issue of gate orphan(s).

## 4. Proposed QDI BCLA and QDI BCLARC

### 4.1. CCLA and BCLA – A Brief Comparison

In general, an N-bit CCLA is constructed by cascading (N/M) M-bit CCLAs where N modulo M equals 0 [30]. The M carry outputs of a M-bit CCLA are produced by lookahead based on the corresponding generate and propagate functions and the carry input. Of the M carry outputs, excepting the most significant lookahead carry output, the remaining (M–1) carry outputs are XOR-ed with the corresponding propagate functions to produce the respective sum output bits. The most significant lookahead carry output produced by a M-bit CCLA is propagated to the next M-bit CCLA to serve as its carry input, which is utilized to produce the corresponding sum output bits of the next stage M-bit CCLA and yield its carry output.

An N-bit BCLA [31], also called the section-carry based CLA [32], is also realized using (N/M) M-bit BCLAs where N modulo M equals 0. However, a M-bit BCLA comprises a M-bit BCLG, three full adders, and a final 3-input XOR function. The M-bit BCLG produces just one carry output by lookahead based on the propagate and generate functions and the carry input, which is then propagated to the successive M-bit BCLA to serve as its carry input. The carry input to an M-bit BCLA along with its corresponding augend and addend inputs are processed by a kind of sub-RCA which is also of size M-bits that features a cascade of full adders and a final 3-input XOR function to produce the respective sum output bits. Hence, the intermediate carries in a M-bit BCLA are not produced by lookahead but rather they are produced in a ripple-carry fashion.

## 4.2. QDI BCLA and BCLARC Architectures

The architectures of the two QDI BCLAs for an example 32-bit addition are shown in Fig. 3. We consider the 32-bit addition here to facilitate a straightforward comparison with the published literature [14] [15].

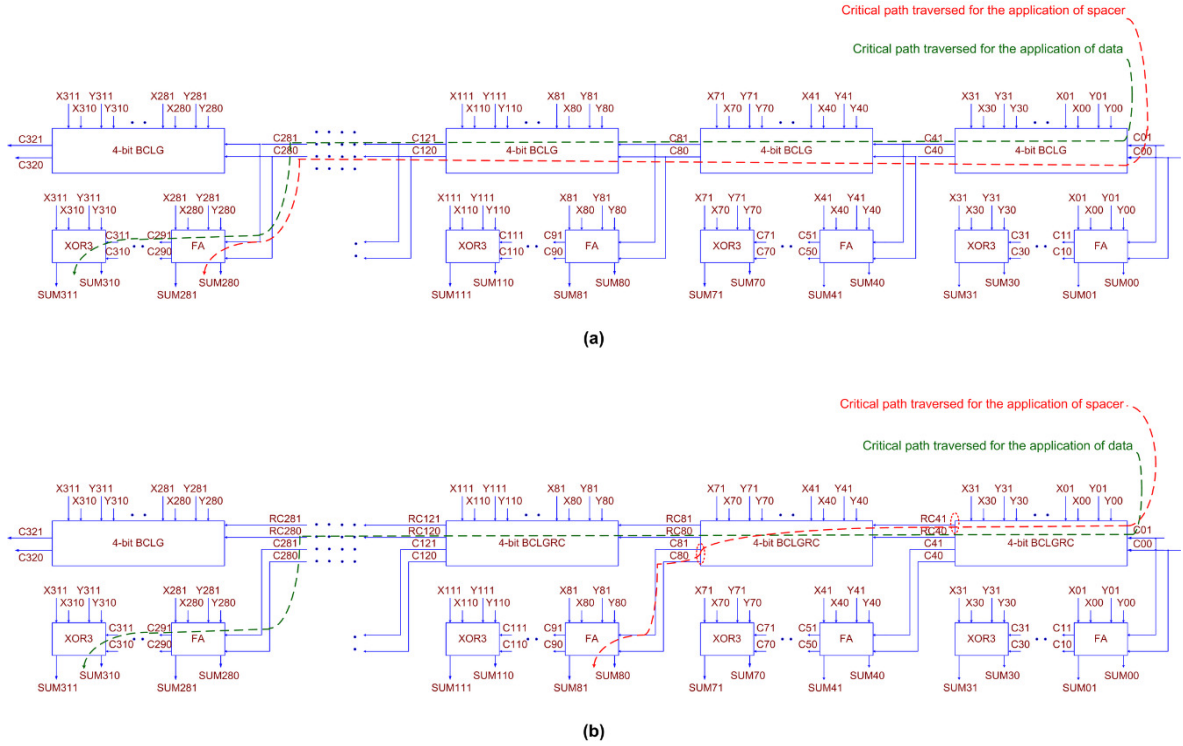

**(a)**

**(b)**

**Fig 3. (a) 32-bit QDI BCLA, and (b) 32-bit QDI BCLARC. The architectures remain the same for RTZ and RTO handshaking. The critical paths traversed for the application of data and spacer also remain the same for RTZ and RTO handshaking. One non-redundant lookahead carry output is produced by each 4-bit QDI BCLG in (a), whereas a non-redundant lookahead carry output and a redundant lookahead carry output is produced by each 4-bit QDI BCLGRC in (b). FA refers to the full adder and XOR3 refers to the 3-input XOR function, and both these belong to (QDI) early output type.**

Fig. 3a shows a 32-bit BCLA that comprises eight 4-bit BCLGs, 24 full adders, and eight 3-input XOR (XOR3) functions. Fig. 3b shows a 32-bit BCLARC that comprises the most significant 4-bit BCLG, seven less significant 4-bit BCLGRCs, 24 full adders and eight XOR3 functions. In Figs. 3a and 3b, (X01, X00) and (Y01, Y00) denote the least significant dual-rail encoded augend and addend inputs, and (X311, X310) and (Y311, Y310) represent the most significant dual-rail encoded augend and addend inputs. The dual-rail encoded carry input and output are denoted by (C01, C00) and (C321, C320) respectively, and the carry input can be set to 0 in the case of RTZ handshaking and set to 1 for RTO handshaking. The critical datapaths traversed for the application of data and spacer in the adders are highlighted by the green and red dashed lines in Figs. 3a and 3b respectively. It can be noticed in Fig. 3 that the 4-bit BCLG, the 4-bit BCLGRC, the full adder, and the XOR3 function form the basic building blocks of the QDI BCLA and the QDI BCLARC.

This work presents the novel and efficient design of a 4-bit BCLG and BCLGRC, which are QDI. The 4-bit BCLG and BCLGRC form the heart of the 4-bit BCLA and the 4-bit BCLARC, which eventually form the building blocks for the QDI BCLA and the QDI BCLARC. QDI realizations of the full adder and the XOR3 function, which were discussed in our previous work [14], have been utilized here to realize the BCLA and the BCLARC.

Gate-level realizations of the 4-bit QDI BCLG/BCLGRC, the early output QDI full adder, and the early output QDI XOR3 function corresponding to RTZ handshaking are shown in Figs. 4a, 4b and 4c respectively. The equivalent gate-level circuits corresponding to RTO handshaking are depicted by Figs. 5a, 5b and 5c respectively. It is proved in [33] that any asynchronous circuit corresponding to RTZ handshaking can be transformed into that corresponding to RTO handshaking and vice-versa by replacing the logic gates by their respective duals while retaining the C-elements and their respective inputs as such. We shall describe the basic building blocks shown in Fig. 4 which correspond to RTZ handshaking, and the discussion will be applicable to those in Fig. 5, which correspond to RTO handshaking.

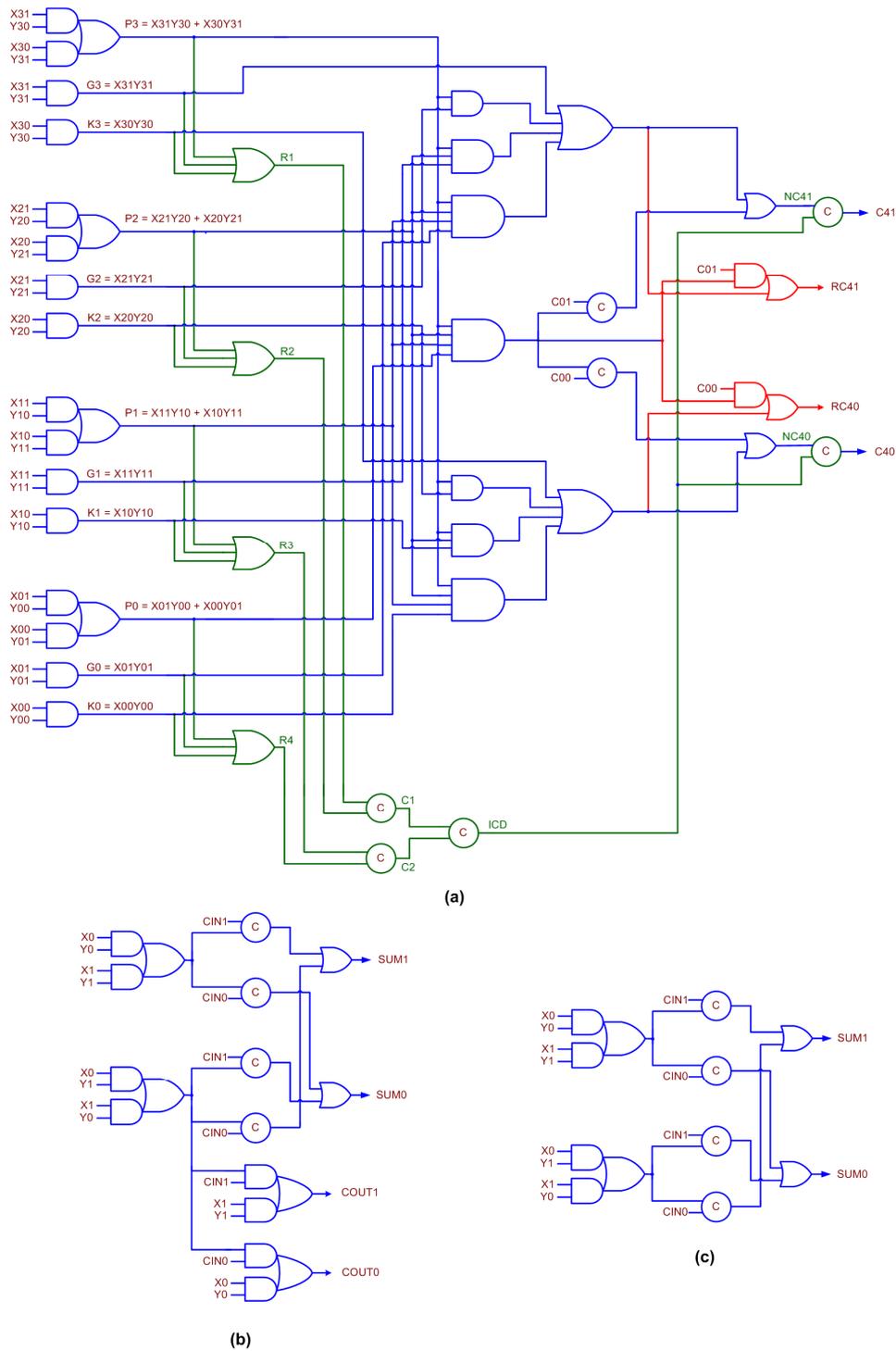

**Fig 4. (a) Proposed QDI 4-bit BCLG/BCLGRC, (b) early output QDI full adder, and (c) early output QDI XOR3 function. All the circuits correspond to 4-phase RTZ handshaking. Note that if the circuit portion shown in red is omitted in (a), it is called 4-bit BCLG; if the circuit portion shown in red is included in (a), it is called 4-bit BCLGRC**

– this interpretation of 4-bit BCLG and 4-bit BCLGRC is also applicable to Fig. 5(a). The circuit portion shown in green lines signifies internal completion detection.

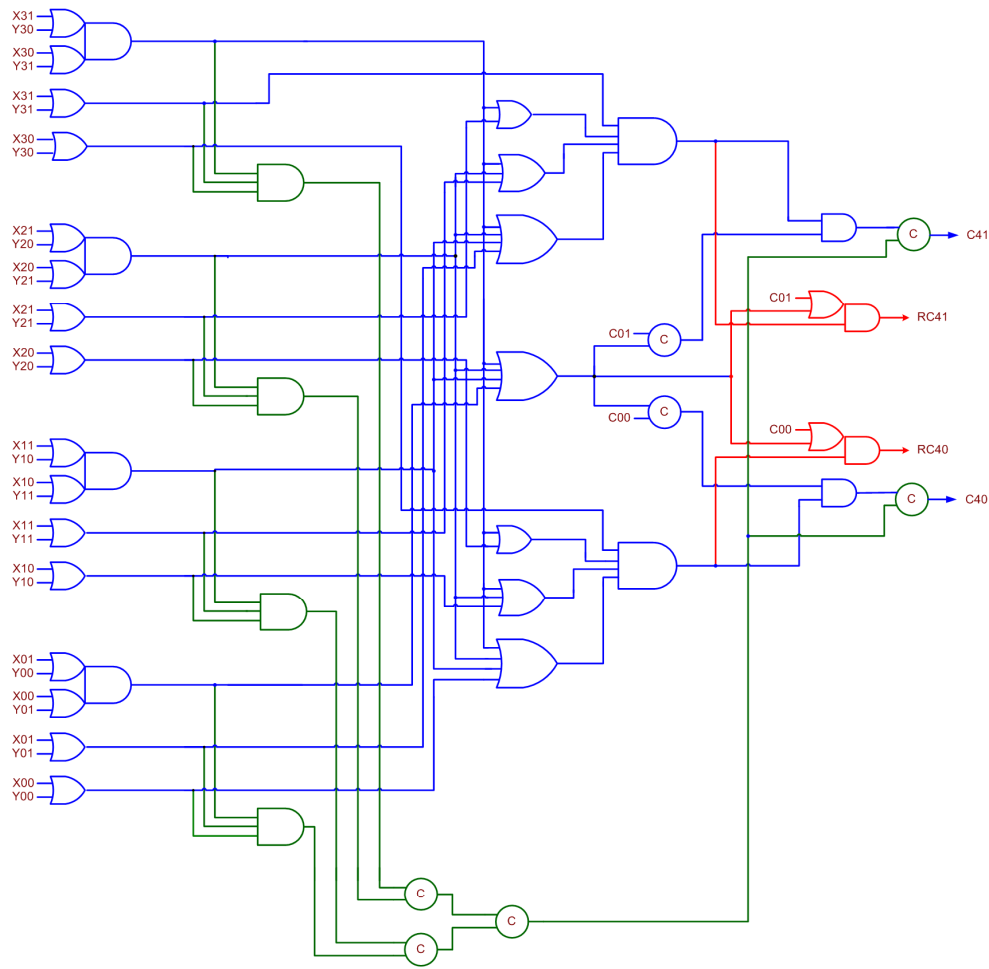

(a)

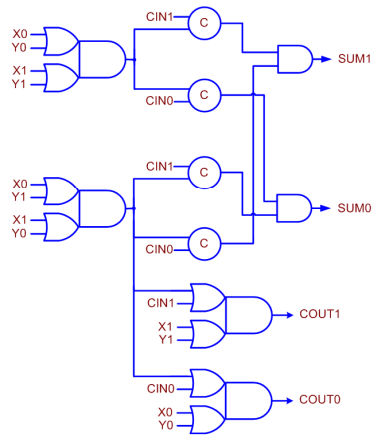

(b)

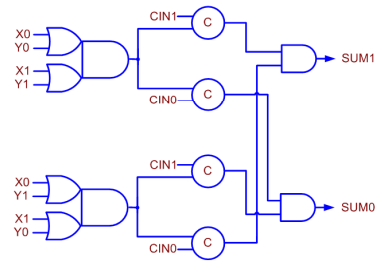

(c)

Fig 5. (a) Proposed QDI 4-bit BCLG/BCLGRC, (b) early output QDI full adder, and (c) early output QDI XOR3 function. All the circuits correspond to 4-phase RTO handshaking. The circuit portion shown in green lines signifies internal completion

**detection, which is crucial to prevent the occurrence of any gate orphan(s) in the 4-bit BCLG or the 4-bit BCLGRC.**

Fig. 4a shows the proposed 4-bit QDI BCLG/BCLGRC. (C41, C40) represent the dual-rail lookahead carry output, with (C01, C00) being the dual-rail carry input, and (RC41, RC40) is the *redundant* dual-rail lookahead carry output, which is logically equivalent to (C41, C40) and the equations for (C41, C40) are given by (1) and (2).

$$C41 = G3 + P3G2 + P3P2G1 + P3P2P1G0 + P3P2P1P0C01 \qquad (1)$$

$$C40 = K3 + P3K2 + P3P2K1 + P3P2P1K0 + P3P2P1P0C00 \qquad (2)$$

In (1) and (2), G3 to G0 represent the carry-generate functions, P3 to P0 represent the carry-propagate functions, and K3 to K0 represent the carry-kill functions. The logic expressions of these functions are given in Fig. 4a. The carry-propagate, carry-generate, and carry-kill functions are mutually orthogonal, which implies that only one of these functions corresponding to a set of primary inputs will be activated for the application of an input data. For example, referring to Fig. 4a, either G3 or P3 or K3 will alone assume 1 during a data phase and the rest will continue to maintain 0 from the earlier RTZ phase. Equations (1) and (2) are thus inherently in the disjoint sum-of-products form.

Note that in Figs. 4a and 5a, if the circuit portion shown in red is omitted, they represent the '4-bit QDI BCLG', and if the circuit portion shown in red is included, they represent the '4-bit QDI BCLGRC'. The circuit portion shown in green lines in Figs. 4a and 5a signifies the internal completion detection, which is crucial to ensure the gate-orphan freedom. The QDI BCLG features only the lookahead carry output (C41, C40), and the QDI BCLGRC features the extra redundant lookahead carry output (RC41, RC40). The proposed 4-bit BCLG and 4-bit BCLGRC belong to the early output type; the BCLG and the BCLGRC will wait for the arrival of the required primary input data to produce the corresponding primary output data.

However, after receiving the spacer on a subset of the primary inputs, all the primary outputs could assume the spacer.

In Fig. 4a, R1, R2, R3, R4, C1, C2, ICD, NC41 and NC40 represent the intermediate outputs. These internal outputs are manifested in Fig. 5a as well. Each set of the respective carry-generate, carry-propagate and carry-kill functions (for example, G3, P3 and K3) are OR-ed in Fig. 4a (AND-ed in Fig. 5a) and their outputs viz. R1 to R4 are given to a C-element tree. The output of the C-element tree is denoted as ICD, which is the output of the internal completion detector. NC41 and NC40 are equivalent to C41 and C40. But NC41 and NC40 are synchronized with ICD to produce C41 and C40. This is to ensure that when C41 and C40 are produced all the internal data processing within the 4-bit BCLG/BCLGRC is completed and all the internal outputs have settled to the correct steady-state. Ensuring internal completion detection is necessary for the proposed BCLG/BCLGRC to guarantee that they are QDI.

To illustrate the importance of and the need for internal completion detection in Fig. 4a (and Fig. 5a), let us assume that P3 = P2 = P1 = G0 = 1 after an RTZ phase. As a result, NC41 would assume 1. Also, R1 = R2 = R3 = R4 = 1. Therefore, C1 = C2 = 1 and ICD = 1. Since NC41 = ICD = 1, C41 = 1 and C40 = 0. Subsequently, in the next RTZ phase, let us assume that only P3, P2 and P1 have become 0 and G0 is still 1. Given this, NC41 will assume 0. Supposing, NC41 was used to represent C41, this will incorrectly convey that the BCLG/BCLGRC has assumed the spacer although the internal data processing has not been completed because G0 has not yet become 0. This violates the QDI principle because in a QDI circuit, the production of the primary outputs should unambiguously confirm the receipt of the primary inputs and the completion of internal computation within the circuit after the processing of data and spacer. This will avoid the likelihood of any problem due to gate orphan(s), which would occur when the output(s) of intermediate gate(s) remain unacknowledged.

## 4.3. Cycle Time Calculation of Proposed QDI BCLA and BCLARC

It would be useful to analyse the (worst-case) cycle times of the proposed QDI BCLA and QDI BCLARC to gain an insight into which of these architectures would be more beneficial in terms of the speed prior to physical realization. To estimate the CT, the estimation of forward and reverse latencies is essential since CT is the summation of forward and reverse latencies.

### 4.3.1. Cycle Time of QDI BCLA

To theoretically estimate the (worst-case) CT of the proposed QDI BCLA that corresponds to RTZ handshaking, let us consider Fig. 3a and Fig. 4. Let $T_{BCLG}$, $T_{FA}$ and $T_{XOR3}$ represent the propagation delays of the QDI early output 4-bit BCLG, the full adder, and the XOR3 function respectively, which are shown in Fig. 4. Also, let the propagation delays of the least significant 4-bit BCLG and the intermediate 4-bit BCLG be denoted as $T_{BCLG\_LS}$ and $T_{BCLG\_INT}$. Given these, the forward latency of the 32-bit QDI BCLA ($FL_{BCLA\_RTZ}$) shown in Fig. 3a that corresponds to RTZ handshaking, which incorporates the QDI building blocks of Fig. 4, is expressed by (3). In (3), the last term on the right-side represents the propagation delay of the input register ($T_{Register}$), which is the propagation delay of the 2-input C-element since the C-element serves as the register.

$$FL_{BCLA\_RTZ} = T_{XOR3} + 3T_{FA} + 6T_{BCLG\_INT} + T_{BCLG\_LS} + T_{Register} \qquad (3)$$

Referring to Fig. 4a, the longest (critical) datapath is traversed in the least significant BCLG which involves an AO22 complex gate, a 3-input OR gate, and three 2-input C-elements. As in the previous works, the 2-input C-element was custom-realized based on a 32/28nm CMOS technology [34] by modifying the AO222 complex gate realization by introducing feedback which required 12 transistors. Besides the C-element, all the other gates in the cell library [34] have been directly utilized. In the subsequent intermediate BCLGs, the datapath traversal would encounter relatively fewer gates which involves a 2-input C-element, a 2-input OR gate, and a final 2-input C-element. The datapath traversal in the full adder would involve an AO22 gate, and the datapath traversal via the XOR3 function would involve a 2-input C-element and a 2-input OR gate.

With $T_{AO22}$, $T_{OR3}$, $T_{CE2}$ and $T_{OR2}$ representing the propagation delays of an AO22 complex gate, a 3-input OR gate, a 2-input C-element, and a 2-input OR gate respectively, (3) is expanded and given by (4). Note that there is a one-to-one correspondence between the terms present on the right-side of (3) and (4).

$$FL_{BCLA\_RTZ} = (T_{CE2} + T_{OR2}) + 3T_{AO22} + 6(2T_{CE2} + T_{OR2}) + (T_{AO22} + T_{OR3} + 3T_{CE2}) + T_{CE2} \quad (4)$$

Let the reverse latency of the QDI BCLA shown in Fig. 3a that corresponds to RTZ handshaking be denoted as $RL_{BCLA\_RTZ}$ which is expressed by (5).

$$RL_{BCLA\_RTZ} = T_{FA} + 6T_{BCLG\_INT} + T_{BCLG\_LS} + T_{Register} \quad (5)$$

Compared to (3), the processing of the spacer in the QDI BCLA involves fewer gates, i.e., two full adders and one XOR3 less. This is because the least significant full adder present in the most significant 4-bit BCLA of Fig. 3a would wait for the arrival of the carry input (C281, C280) to process it to produce the sum output bit (SUM281, SUM280). Referring to Fig. 4b, the carry outputs of all the full adders can be produced early and when they are given as the carry inputs for the successive full adders in the cascade, the sum outputs of the full adders could be produced simultaneously. This time delay is less compared to the reverse latency of the QDI BCLA shown in Fig. 3a. Thus, (5) is expanded and given as (6), and there is a one-to-one correspondence between the terms present on the right-side of (5) and (6).

$$RL_{BCLA\_RTZ} = (T_{CE2} + T_{OR2}) + 6(2T_{CE2} + T_{OR2}) + (T_{AO22} + T_{OR3} + 3T_{CE2}) + T_{CE2} \quad (6)$$

The CT of the QDI BCLA (Fig. 3a) can be calculated by substituting the propagation delays of the minimum-size gates present in the cell library in (4) and (6), and then adding up the forward and reverse latencies. Based on the theoretical calculations, the forward and reverse latencies of the 32-bit QDI BCLA are found to be 2.583ns and 2.367ns, resulting in a CT of 4.95ns for RTZ handshaking.

The detailed expressions for forward and reverse latencies corresponding to RTO handshaking are given by (7) and (8) with reference to Fig. 3a and Fig. 5. Equations (7) and (8) are deduced by replacing the propagation delays of the gates mentioned in (4) and (6) with the propagation delays of their dual gates, however, with the exception of $T_{CE2}$, which is retained as such. This is because the 2-input C-elements and their respective inputs are retained while transforming a circuit corresponding to RTZ handshaking into that that corresponds to RTO handshaking, according to the logic transformation rules given in [33].

$$FL_{BCLA\_RTO} = (T_{CE2} + T_{AND2}) + 3T_{OA22} + 6(2T_{CE2} + T_{AND2}) + (T_{OA22} + T_{AND3} + 3T_{CE2}) + T_{CE2}$$

$$(7)$$

$$RL_{BCLA\_RTO} = (T_{CE2} + T_{AND2}) + 6(2T_{CE2} + T_{AND2}) + (T_{OA22} + T_{AND3} + 3T_{CE2}) + T_{CE2} \qquad (8)$$

Based on (7) and (8), the forward and reverse latencies of the 32-bit QDI BCLA, shown in Fig. 3a, are calculated to be 2.842ns and 2.632ns, resulting in a CT of 5.474ns for RTO handshaking.

### 4.3.2. Cycle Time of QDI BCLARC

To theoretically estimate the (worst-case) CT of the proposed QDI BCLARC that corresponds to RTZ handshaking, let us consider Fig. 3b and Fig. 4. In Fig. 3b, one most significant 4-bit BCLA and seven less significant 4-bit BCLARCs are used. The use of the 4-bit BCLA for the most significant adder nibble position is because only one (non-redundant) lookahead carry output should be produced which represents the carry overflow.

Starting from the least significant 4-bit BCLARC, each 4-bit BCLARC produces a non-redundant lookahead carry output and a redundant lookahead carry output. The redundant lookahead carry output of a 4-bit BCLGRC is propagated to the successive 4-bit BCLGRC (or 4-bit BCLG) as its carry input, whereas the non-redundant lookahead carry output is propagated to a cascade of three full adders and an XOR3 present in the successive 4-bit BCLARC (or 4-bit BCLA).

Referring to Fig. 4a, the longest (critical) datapath would be traversed in the least significant 4-bit BCLGRC involving an AO22 complex gate, a 4-input AND gate, a 4-input OR gate, and an AO21 complex gate. In the subsequent intermediate 4-bit BCLGRCs, the datapath traversal would involve just one AO21 complex gate.

The forward latency of the BCLARC corresponding to RTZ handshaking ($FL_{BCLARC\_RTZ}$), which is shown in Fig. 3b, is expressed by (9), where $T_{BCLGRC}$ denotes the propagation delay of the 4-bit BCLGRC shown in Fig. 4a. In (9), $T_{BCLGRC\_INT}$ specifies the propagation delay of a 4-bit BCLGRC present in an intermediate nibble position of the adder, and $T_{BCLGRC\_LS}$ specifies the propagation delay of the least significant 4-bit BCLGRC.

$$FL_{BCLARC\_RTZ} = T_{XOR3} + 3T_{FA} + 6T_{BCLGRC\_INT} + T_{BCLGRC\_LS} + T_{Register} \qquad (9)$$

Equation (9) is expanded and given by (10), where $T_{AO21}$, $T_{AND4}$ and $T_{OR4}$ denote the propagation delays of the AO21 complex gate, the 4-input AND gate, and the 4-input OR gate respectively. There is a one-to-one correspondence between the terms present on the right-side of (9) and (10).

$$FL_{BCLARC\_RTZ} = (T_{CE2} + T_{OR2}) + 3T_{AO22} + 6T_{AO21} + (T_{AO22} + T_{AND4} + T_{OR4} + T_{AO21}) + T_{CE2} \quad (10)$$

The critical datapath traversed for the application of the spacer in the case of the 32-bit QDI BCLARC is highlighted by the red dashed line in Fig. 3b. Since the 4-bit QDI BCLGRC shown in Fig. 4a is of early output type, and because this is used to construct the QDI BCLARC of Fig. 3b, the redundant lookahead carry outputs of all the 4-bit BCLGRCs could assume the spacer simultaneously. But, the redundant lookahead carry output produced by a 4-bit BCLGRC is given as the carry input for the successive 4-bit BCLGRC (or 4-bit BCLG) to produce the corresponding non-redundant lookahead carry output. This carry output then serves as the carry input for the least significant full adder present in the following 4-bit BCLARC (or 4-bit BCLA) to produce the corresponding sum output bit.

With $RL_{BCLARC\_RTZ}$ representing the reverse latency of the QDI BCLARC that corresponds to RTZ handshaking, as shown in Fig. 3b, and also referring to Fig. 5, it is expressed by (11). In (11), $T_{BCLG\_LS}$ may be replaced by $T_{BCLG\_INT}$ without any loss of generality since the reverse latency would be the same. The expanded version of (11) is given by (12), and there exists a one-to-one correspondence between the terms present on the right-side of (11) and (12).

$$RL_{BCLARC\_RTZ} = T_{FA} + T_{BCLG\_INT} + T_{BCLG\_LS} + T_{Register} \qquad (11)$$

$$RL_{BCLARC\_RTZ} = (T_{CE2} + T_{OR2}) + (2T_{CE2} + T_{OR2}) + (T_{AO22} + T_{AND4} + T_{OR4} + T_{AO21}) + T_{CE2} \quad (12)$$

Based on (10) and (12), the forward and reverse latencies of the QDI BCLARC, shown in Fig. 3b, corresponding to RTZ handshaking are calculated as 1.171ns and 0.849ns, which results in a CT of 2.02ns.

The detailed expressions for forward and reverse latencies corresponding to RTO handshaking are given by (13) and (14). Equations (13) and (14) are deduced by replacing the propagation delays of the gates mentioned in (10) and (12) with the propagation delays of their dual gates, however, excluding $T_{CE2}$ which is alone retained as such.

$$FL_{BCLARC\_RTO} = (T_{CE2} + T_{AND2}) + 3T_{OA22} + 6T_{OA21} + (T_{OA22} + T_{OR4} + T_{AND4} + T_{OA21}) + T_{CE2}$$
$$(13)$$

$$RL_{BCLARC\_RTO} = (T_{CE2} + T_{AND2}) + (2T_{CE2} + T_{AND2}) + (T_{OA22} + T_{OR4} + T_{AND4} + T_{OA21}) + T_{CE2} \quad (14)$$

Based on (13) and (14), the forward and reverse latencies of the 32-bit QDI BCLARC (Fig. 3b) corresponding to RTO handshaking are calculated as 1.245ns and 0.933ns, which results in a CT of 2.178ns.

Based on the theoretical calculations of CT, it is noted that the QDI BCLARC architecture achieves 59.1% and 60.2% reductions in the CT than the QDI BCLA architecture for the 32-bit addition with respect to RTZ and RTO handshaking respectively. This implies the former architecture (BCLARC) is more beneficial than the latter for performing addition at an enhanced speed. Based on the simulation results obtained, which will be discussed in the next section, it is found that the QDI BCLARC architecture achieves 57% and 55.7% reductions in CT over the QDI BCLA architecture for a 32-bit addition with respect to RTZ and RTO handshaking respectively. Thus, a good correlation is evident between the theoretical calculations and the practical estimations of CT. Thus, although the theoretical calculations may be quite approximate, nevertheless they are useful because they give a correct and valuable design insight, which is that the QDI BCLARC architecture is preferable to the QDI BCLA architecture. Nevertheless, differences between the theoretical calculations and the practical estimations are expected because the interconnect delays and the parasitic are not accounted for in the theoretical calculations.

## 5. Results and Discussion

A total of fifty-six 32-bit QDI and non-QDI (relative-timed) asynchronous adders, which correspond to various architectures such as RCA, CSLA, CCLA, BCLA, BCLARC, and hybrid BCLARC-RCA were physically realized using a 32/28nm CMOS technology [34]. Of the fifty-six asynchronous adders, twenty-eight of them correspond to the RTZ handshaking and a similar number corresponds to the RTO handshaking. As mentioned earlier, the 2-input C-element was custom-realized by modifying the AO222 gate to implement the asynchronous adders. A typical-case PVT specification of the high $V_t$ standard digital cell library with a recommended supply voltage of 1.05V and an operating junction temperature of 25°C was considered for the physical implementations and simulations. The registers and completion detectors associated with the asynchronous adders are maintained the same with respect to RTZ and RTO handshaking, separately. This implies the differences between the simulation results of the adders are entirely attributable to the differences between their logic compositions.

About 2000 (random) input vectors including data and spacer, which separately correspond to RTZ and RTO handshaking, as used in our prior work [14], were used to verify the functionalities of the adders. The input vectors corresponding to RTZ and RTO handshaking bear a logical equivalence. Functional simulations of all the adders were successfully performed and their respective switching activities were captured which were subsequently used to estimate the average power dissipation. Synopsys EDA tools were used to estimate the design metrics of the adders.

The design metrics estimated include forward and reverse latencies, CT, area, and average power dissipation. The forward latency of an asynchronous circuit is similar to the critical path delay of a synchronous circuit and hence it is directly estimated. The reverse latency may differ from the forward latency, which is evident from the critical datapaths highlighted in Figs. 3a and 3b. The reverse latencies of the asynchronous adders were estimated from the gate-level simulation timing data, and this method was followed for RTZ and RTO handshaking as adopted in our previous work [14].

**Table 1. Design metrics of several 32-bit asynchronous adders (QDI and non-QDI) corresponding to RTZ handshaking.**

| Adder Architecture | Adder Legends | Literature Reference(s) | FL[1] (ns) | RL[2] (ns) | CT (ns) | Area ($\mu m^2$) | Power ($\mu W$) |
|---|---|---|---|---|---|---|---|
| RCA | Z1 | [36] | 14.61 | 14.61 | 29.22 | 2529.00 | 2190 |
| | Z2 | [37][3] | 9.26 | 9.26 | 18.52 | 2504.60 | 2181 |
| | Z3 | [25] | 9.04 | 9.04 | 18.08 | 2293.14 | 2172 |
| | Z4 | [37][4] | 8.24 | 8.24 | 16.48 | 2423.27 | 2177 |
| | Z5 | [38] | 7.00 | 7.00 | 14.00 | 2016.63 | 2171 |
| | Z6 | [39] | 4.43 | **0.58** | 5.01 | 2097.96 | 2174 |
| | Z7 | [23] | 3.32 | 0.73 | 4.05 | 2049.16 | 2171 |
| | Z8 | [24] | 3.10 | 0.61 | 3.71 | 1658.80 | 2161 |
| | Z9 | [15][5] | 2.91 | 0.62 | 3.53 | 1658.80 | 2168 |
| Uniform CSLA | Z10 | | 2.46 | 1.89 | 4.35 | 3000.17 | 2293 |
| Non-uniform CSLA | Z11 | [40] | 3.23 | 3.23 | 6.46 | 3384.44 | 2312 |
| BCLA | Z12 | | 3.31 | 2.93 | 6.24 | 2951.88 | 2191 |
| BCLARC | Z13 | [39] | 2.46 | 1.69 | 4.15 | 2987.46 | 2192 |
| BCLA | Z14 | | 3.14 | 2.88 | 6.02 | 2915.29 | 2188 |
| BCLARC | Z15 | [42] | 2.32 | 1.68 | 4.00 | 2950.87 | 2189 |
| CCLA | Z16 | [41] | 2.75 | 2.75 | 5.50 | 2569.65 | 2177 |

| Adder Architecture | Adder Legends | Literature Reference(s) | FL (ns) | RL (ns) | CT (ns) | Area (µm²) | Power (µW) |
|---|---|---|---|---|---|---|---|
| BCLA | Z17 | | 3.13 | 2.88 | 6.01 | 2524.92 | 2178 |
| BCLARC | Z18 | [42] | 2.31 | 1.67 | 3.98 | 2560.50 | 2179 |
| BCLA | Z19 | | 2.76 | 2.50 | 5.26 | 2209.78 | 2174 |
| BCLARC | Z20 | | 2.01 | 1.38 | 3.39 | 2245.36 | 2176 |
| Hybrid BCLARC-RCA1 | Z21 | [14] | 1.93 | 1.38 | 3.31 | 2171.41 | 2174 |
| Hybrid BCLARC-RCA1 | Z22 | | 1.97 | 1.38 | 3.35 | 2097.45 | 2172 |
| Hybrid BCLARC-RCA1 | Z23 | | 2.23 | 1.38 | 3.61 | 2023.49 | 2170 |
| BCLA | Z24 | | 3.46 | 3.20 | 6.66 | 2307.37 | 2187 |
| BCLARC | Z25 | | **1.76** | 1.11 | **2.87** | 2342.95 | 2188 |
| Hybrid BCLARC-RCA1 | Z26 | Proposed | 1.86 | 1.11 | 2.97 | 2256.80 | 2184 |
| Hybrid BCLARC-RCA2 | Z27 | | 2.11 | 1.11 | 3.22 | 2170.64 | 2181 |
| Hybrid BCLARC-RCA3 | Z28 | | 2.36 | 1.11 | 3.47 | 2084.49 | 2178 |

[1] Forward Latency; [2] Reverse Latency; [3] Uses strong-indication full adder; [4] Uses weak-indication full adder; [5] Uses LOPT_EO_FA of [15] leading to less CT.

The estimated design metrics of the adders corresponding to RTZ handshaking are given in Table 1, and the design metrics corresponding to RTO handshaking are given in Table 2. The optimum values of forward and reverse latencies and CT are highlighted in bold-face in Tables 1 and 2. Adder legends are provided in the second column of Tables 1 and 2 to conveniently refer to the various adders during the discussion. The related literature references pertaining to those adders are also given in Tables 1 and 2.

**Table 2. Design metrics of several 32-bit asynchronous adders (QDI and non-QDI) corresponding to RTO handshaking.**

| Adder Architecture | Adder Legends | Literature Reference(s) | FL[1] (ns) | RL[2] (ns) | CT (ns) | Area (µm²) | Power (µW) |
|---|---|---|---|---|---|---|---|
| | O1 | [36] | 14.15 | 14.15 | 28.30 | 2529.00 | 2185 |
| | O2 | [37][3] | 8.74 | 8.74 | 17.48 | 2374.48 | 2167 |
| | O3 | [25] | 8.88 | 8.88 | 17.76 | 2293.15 | 2168 |
| | O4 | [37][4] | 8.03 | 8.03 | 16.06 | 2358.21 | 2167 |
| RCA | O5 | [38] | 6.95 | 6.95 | 13.90 | 2016.63 | 2167 |
| | O6 | [39] | 3.79 | **0.56** | 4.35 | 2097.96 | 2170 |
| | O7 | [23] | 3.31 | 0.72 | 4.03 | 2049.16 | 2167 |
| | O8 | [24] | 2.93 | 0.61 | 3.54 | 1658.80 | 2157 |
| | O9 | [15][5] | 2.76 | 0.61 | 3.37 | 1658.80 | 2164 |
| Uniform CSLA | O10 | | 2.38 | 1.85 | 4.23 | 3000.17 | 2285 |
| Non-uniform CSLA | O11 | [40] | 3.15 | 3.08 | 6.23 | 3384.44 | 2303 |
| BCLA | O12 | | 3.19 | 2.86 | 6.05 | 2984.41 | 2184 |
| BCLARC | O13 | [39] | 2.36 | 1.69 | 4.05 | 3019.99 | 2185 |
| BCLA | O14 | | 3.10 | 2.84 | 5.94 | 2947.82 | 2182 |
| BCLARC | O15 | [42] | 2.30 | 1.67 | 3.97 | 2983.40 | 2183 |
| CCLA | O16 | [41] | 2.73 | 2.73 | 5.46 | 2553.39 | 2169 |
| BCLA | O17 | | 3.06 | 2.76 | 5.82 | 2557.45 | 2171 |
| BCLARC | O18 | [42] | 2.26 | 1.66 | 3.92 | 2593.03 | 2172 |
| BCLA | O19 | | 2.73 | 2.50 | 5.23 | 2193.52 | 2167 |
| BCLARC | O20 | | 1.95 | 1.37 | 3.32 | 2229.10 | 2168 |
| Hybrid BCLARC-RCA1 | O21 | [14] | 1.88 | 1.37 | 3.25 | 2157.17 | 2167 |
| Hybrid BCLARC-RCA1 | O22 | | 1.89 | 1.37 | 3.26 | 2085.25 | 2165 |
| Hybrid BCLARC-RCA1 | O23 | | 2.13 | 1.37 | 3.50 | 2013.33 | 2164 |
| BCLA | O24 | | 3.38 | 3.14 | 6.52 | 2315.51 | 2180 |
| BCLARC | O25 | | **1.74** | 1.15 | **2.89** | 2351.09 | 2181 |
| Hybrid BCLARC-RCA1 | O26 | Proposed | 1.78 | 1.15 | 2.93 | 2263.92 | 2178 |
| Hybrid BCLARC-RCA2 | O27 | | 2.02 | 1.15 | 3.17 | 2176.74 | 2175 |
| Hybrid BCLARC-RCA3 | O28 | | 2.26 | 1.15 | 3.41 | 2089.57 | 2172 |

CT governs the speed of a QDI or a relative-timed asynchronous circuit that utilizes delay-insensitive data encoding and a 4-phase handshaking, and PCTP governs the low power/energy aspect. The PCTPs of the asynchronous adders were calculated and normalized. The normalization was performed such that the highest PCTP among the set of asynchronous adders corresponding to a handshake protocol was normalized to 1, and the actual PCTPs of the remaining adders were divided by the highest PCTP. Thus, after normalization, the least value of the PCTP reflects the optimum low power/energy design. The plots of normalized CT and PCTP values corresponding to RTZ handshaking are shown side-by-side in Figs. 6a and 6b, and the similar plots for RTO handshaking are portrayed by Figs. 7a and 7b.

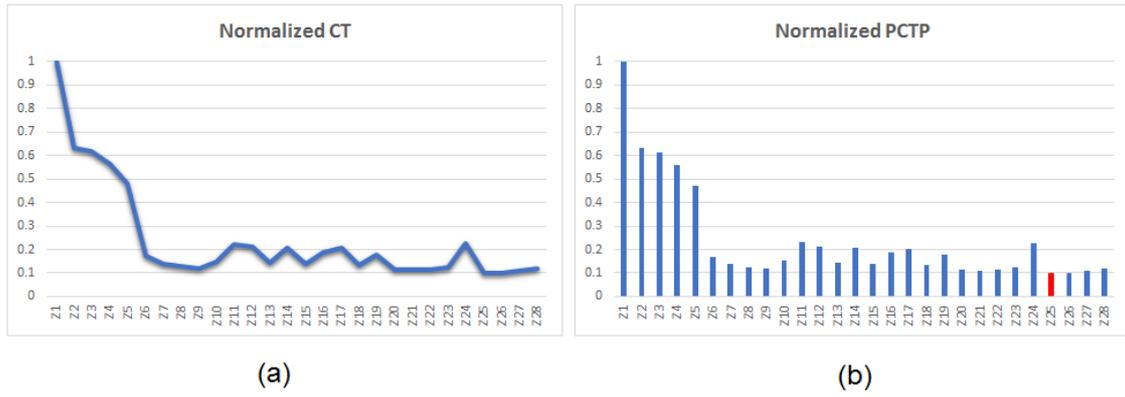

(a)          (b)

**Fig 6. Plots of normalized values of (a) CT and (b) PCTP of several 32-bit asynchronous adders corresponding to RTZ handshaking. The adder legends are referenced in Table 1. The red bar in (b) corresponds to the proposed 32-bit BCLARC which is energy-efficient than the rest.**

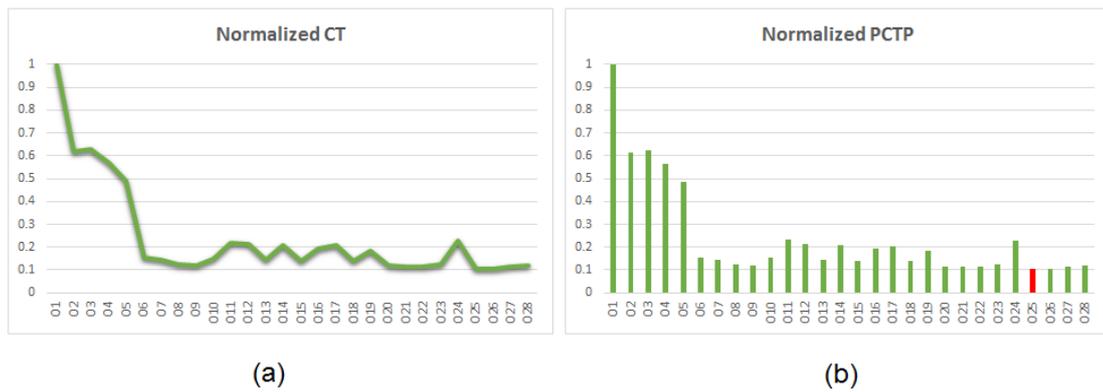

(a)          (b)

**Fig 7. Plots of normalized values of (a) CT and (b) PCTP of several 32-bit asynchronous adders corresponding to RTO handshaking. The adder legends are referenced in Table 2. The red bar in (b) corresponds to the proposed 32-bit BCLARC which is energy-efficient than the rest.**

The common trends noticeable with respect to the design metrics in Table 1 are almost mirrored in the case of Table 2. Hence, some general inferences can be drawn based on Tables 1 and 2. In the case of synchronous design, the RCA architecture is the slowest, while CLA and CSLA architectures are known to be substantially faster. However, in the case of QDI and non-QDI (relative-timed) asynchronous adders, which employ delay-insensitive data encoding and a 4-phase handshaking, the observation made with respect to a synchronous adder design may not hold well. This is because CLA and CSLA architectures could facilitate a significant reduction in the forward latency compared to the RCA architecture. But the RCA architecture can facilitate a substantial reduction in the reverse latency compared to CLA and CSLA architectures. Hence, the RCA may enable a reduced CT than some CLAs and CSLAs. For example, Z9, which reports the least CT among the RCAs in Table 1 has a data-dependent forward latency like any other latency-optimised RCA or CSLA or CLA or CLA-RCA. However, the reverse latency of Z9 is a constant and optimal since only one full adder delay would be encountered for processing the spacer. A comparison of the CT of Z9 with the CTs of Z10 to Z19 and Z24 in Table 1 reveals that the former, which is an RCA reports less CT than the latter which are CSLAs and CLAs. A similar observation can be made upon comparing O9 with O10 to O19 and O24 in Table 2.

In Tables 1 and 2, Z24 to Z28 and O24 to O28 denote the proposed CLAs and hybrid CLA-RCAs corresponding to RTZ and RTO handshaking respectively. Relatively fewer gates would be traversed in the critical data path for processing the data in a CLA or a CSLA compared to the many gates which would be traversed for processing the data in an RCA in its critical data path. For example, referring to (10), to process the data, the critical path traversed in the proposed BCLARC (Z25 of Table 1) would encounter two 2-input C-elements including a register, seven AO21 gates, four AO22 gates, a 4-input AND gate, a 4-input OR gate and a 2-input OR gate, resulting in a practical forward latency of 1.76ns. On the other hand, the critical path traversed in an RCA (say, Z6 of Table 1) to process the data would encounter a register, thirty-one AO222 gates, a 2-input C-element and a 2-input OR gate, resulting in a practical forward latency of 4.43ns, which is 151.7% greater than that of Z25. We considered Z6 for comparison here because it reports the optimum reverse latency in Table 1.

To process the spacer, in the case of Z6, fewer gates would be encountered in an RCA viz. an input register, a AO222 gate, a 2-input C-element and a 2-input OR gate compared to the many gates which would be encountered in the proposed BCLARC (Z25) involving an input register, a AO22 gate, a 4-input AND gate, a 4-input OR gate, a AO21 gate, two 2-input C-elements and two 2-input OR gates. As a result, Z6 reports a 47.7% reduction in the reverse latency compared to Z25. Nevertheless, Z25 reports a 42.7% less CT compared to Z6 because of the significantly reduced forward latency. Z25 has a CT of 2.87ns, which is less than the CTs of all the other adders mentioned in Table 1. Likewise, in Table 2, O25 reports the least CT amongst all the adders.

Among the RCAs mentioned in Tables 1 and 2, Z9 and O9 have the least CT. Z9 and O9 are relative-timed (non-QDI) RCAs and they comprise the latency optimised early output full adder (LOPT_EO_FA) of [15]. The forward latency of Z9 and O9 is data-dependent while their reverse latency is a constant, which is in fact optimal. Compared to Z9 (O9), the proposed BCLARC i.e., Z25 (O25) has 18.7% (14.2%) less CT and 17.9% (13.6%) less PCTP.

CLA architectures which incorporate redundant carries tend to have reduced forward and reverse latencies and CT compared to those of plain CLA architectures which do not have redundant carries, i.e., the QDI BCLARC architecture outperforms the QDI BCLA architecture in terms of the timing. This observation is substantiated by the discussions given in Section IV, and is further evidenced upon comparing Z12 and Z13, Z14 and Z15, Z17 and Z18, Z19 and Z20, and Z24 and Z25 in Table 1, and by comparing O12 and O13, O14 and O15, O17 and O18, O19 and O20, and O24 and O25 in Table 2. It has been noted in [35] that introducing redundant logic, which can be interpreted as the redundant carry logic introduced in the BCLARC architecture, facilitates an improved optimization of the timing parameters.

In the case of CCLAs [41] i.e., Z16 of Table 1 and O16 of Table 2, which are QDI and of early output type, their forward and reverse latencies are equal. This is because the same critical datapath would be traversed for processing the data and the spacer, and the critical data path is data-dependent. Moreover, there is no opportunity for introducing redundant carries in the CCLA architecture to further speed-up the carry propagation since the lookahead carry output of say, a 4-bit CCLA is provided as the carry input for the successive 4-bit CCLA in the cascade. As a result, CTs of Z16 and O16 are considerably greater than the CTs of all the BCLARCs. The proposed BCLARC i.e., Z25 achieves a 47.8% reduction in CT and a 47.6% reduction in PCTP compared to Z16. Based on RTO handshaking, O25 achieves a 47.1% reduction in CT and a 46.8% reduction in PCTP compared to O16.

In Tables 1 and 2, hybrid BCLARC-RCAs are also considered. They are denoted by Z21 to Z23 and Z26 to Z28 in Table 1, and O21 to O23 and O26 to O28 in Table 2. A hybrid BCLARC-RCA architecture replaces one or more less significant sub-BCLARC(s) with a similar size RCA, which is composed using full adders. For example, Z21 and Z26, Z22 and Z27, and Z23 and Z28 in Table 1 incorporate a 4-bit RCA, an 8-bit RCA and a 12-bit RCA in the least significant adder bit positions as a corresponding replacement for one, two and three instances of a 4-bit BCLARC respectively. While the replacement of one or more 4-bit BCLARCs by a corresponding size RCA could help to reduce the area, it is not guaranteed that such a replacement will have a beneficial impact on the CT, and rather the contrary might result.

The CTs of Z21, Z22 and Z23, and Z26, Z27 and Z28 given in Table 1 reveal that increasing the size of the sub-RCA when used in the least significant adder bit positions increases the forward latency of the hybrid BCLARC-RCAs although their reverse latencies remain a constant. The constant reverse latency is because of the traversal of the same critical datapath, shown using the red dashed line in Fig. 3b. The forward latencies of Z26, Z27 and Z28, belonging to Table 1, are expressed by (15) to (17). These are obtained by modifying (10) while considering the replacement of sub-BCLARC(s) with a similar sized sub-RCA. To construct the sub-RCA, the QDI early output full adder of [24] was used, and this was used to construct the hybrid BCLARC-RCAs in [14] as well.

$$FL_{Z26} = (T_{CE2} + T_{OR2}) + 3T_{AO22} + 6T_{AO21} + 5T_{AO22} + T_{CE2} \qquad (15)$$

$$FL_{Z27} = (T_{CE2} + T_{OR2}) + 3T_{AO22} + 5T_{AO21} + 9T_{AO22} + T_{CE2} \qquad (16)$$

$$FL_{Z28} = (T_{CE2} + T_{OR2}) + 3T_{AO22} + 4T_{AO21} + 13T_{AO22} + T_{CE2} \qquad (17)$$

By substituting the propagation delays of the gates in (15), (16) and (17), the theoretical forward latencies of Z26, Z27 and Z28 in Table 1 were calculated as 1.226ns, 1.451ns and 1.676ns respectively. In Section IV (C2), the theoretical forward latency of Z25 was calculated as 1.171ns. Hence, theoretically, Z25 has a reduced forward latency than Z26, Z27 and Z28, which is supported by the practical estimates given in Table 1. This once again validates the accuracy of our theoretical delay modelling.

The reverse latency of Z25 was theoretically calculated as 0.849ns in Section IV (C2), and the same reverse latency is applicable for Z26, Z27 and Z28 in Table 1. Hence, theoretically, the CTs of Z25, Z26, Z27 and Z28 equate to 2.02ns, 2.075ns, 2.3ns and 2.525ns respectively. This shows that Z25, which is the proposed BCLARC, has a reduced CT than the CTs of Z26, Z27 and Z28, which are the hybrid BCLARC-RCAs. Theoretically, the CT of Z25 is 2.7% less than the CT of Z26, and practically (based on the results given in Table 1), the CT of Z25 is found to be 3.4% less than the CT of Z26. Again, there is a correlation between the theoretical calculations and the practical estimates of CT, and the theoretical calculations provide a valuable design insight.

Based on (15), (16) and (17), and considering the duals of the respective gates except for the 2-input C-elements, the forward latencies of O26, O27 and O28, which are the RTO counterparts of Z26, Z27 and Z28, mentioned in Table 2, can also be theoretically modelled. This can be done by modifying (15), (16) and (17) by replacing the propagation delays of the gates mentioned with the propagation delays of their dual gates, however, excluding the delay of the 2-input C-element alone which is retained as such. Theoretically, the forward latencies of O26, O27 and O28 are calculated as 1.286ns, 1.497ns and 1.708ns respectively. Given that (14) is applicable for O25, O26, O27 and O28, their CTs are theoretically calculated as 2.178ns, 2.219ns, 2.43ns and 2.641ns. This shows that O25, which represents the RTO equivalent of the proposed BCLARC, has a reduced CT than the CTs of hybrid BCLARC-RCAs viz. O26, O27 and O28. Hence, based on the proposed 4-bit BCLGRCs, portrayed by Figs. 4a and 5a, it is inferred that the proposed BCLARC is preferable to hybrid BCLARC-RCAs based on CT with respect to RTZ and RTO handshaking.

The proposed BCLARC achieves a substantial reduction in the CT compared to the CTs of other BCLARCs and in comparison with the optimum CT of a hybrid BCLARC-RCA, as reported in a latest work [14]. Hence, hybrid BCLARC-RCAs corresponding to [42] were not considered as they are anyways sub-optimum.

In terms of the area, the RCA architecture is preferable to CLA and CSLA architectures, and this observation is valid for synchronous as well as asynchronous adders. With respect to power dissipation, almost all the asynchronous adders, whether they are QDI or non-QDI, dissipate quite nearly the same power with the standard deviation from the mean of the power dissipation estimated to be 33.5 for RTZ handshaking and 33.1 for RTO handshaking. The small values of standard deviations are because the asynchronous adders mentioned in Tables 1 and 2 embed the monotonic cover constraint, discussed in Section III (B). Hence, the power dissipation of various QDI and non-QDI (relative-timed) adders do not vary considerably and are confined to small ranges of 2161μW – 2312μW in the case of Table 1, and 2157μW – 2303μW in the case of Table 2.

Given that the average power dissipation of several asynchronous adders is quite nearly the same, it may be observed that the differences in their PCTP are mainly due to the differences in their CTs with respect to RTZ and RTO handshaking. This may be evident upon perusing Figs. 6a and 6b, and Figs. 7a and 7b.

## 6. Conclusions

This article presented the design of a new asynchronous QDI early output sub-BCLG/BCLGRC that forms the basis for constructing a QDI early output BCLA/BCLARC. In specific, we discussed the design of a 4-bit QDI BCLA and a 4-bit QDI BCLARC which serve as the building blocks for constructing the QDI early output BCLARC. For an example, we considered the 32-bit addition and compared the proposed QDI BCLARC with several asynchronous adders, which are QDI and non-QDI (relative-timed). Further, hybrid BCLARC-RCAs were considered for the comparison. The simulation results show that the proposed QDI early output BCLARC (Z25 of Table 1 and O25 of Table 2) is efficient in terms of speed (CT) as well as low power/energy (PCTP).

With respect to RTZ handshaking, the proposed QDI BCLARC (Z25 of Table 1) achieved the following reductions in design metrics over its counterparts for 32-bit addition: i) 22.6% and 21.7% reductions in CT and PCTP respectively compared to an optimum QDI early output RCA (i.e., Z8), ii) 18.7% and 17.9% reductions in CT and PCTP respectively compared to an optimum relative-timed RCA (i.e., Z9), iii) 34% and 37% reductions in CT and PCTP respectively compared to an optimum uniform input-partitioned QDI early output CSLA (i.e., Z10), iv) 47.8% and 47.6% reductions in CT and PCTP respectively compared to an optimum QDI early output CCLA (i.e., Z16), v) 45.4% and 45.1% reductions in CT and PCTP respectively compared to an optimum QDI early output BCLA (i.e., Z19), vi) 15.3% and 14.9% reductions in CT and PCTP respectively compared to an optimum QDI early output BCLARC (i.e., Z20), and vii) 13.3% and 12.7% reductions in CT and PCTP respectively compared to an optimum QDI early output hybrid BCLARC-RCA (i.e., Z21).

Based on RTO handshaking, the proposed QDI BCLARC (O25 of Table 2) achieved the following reductions in design metrics over its counterparts for 32-bit addition: i) 18.4% and 17.5% reductions in CT and PCTP respectively compared to an optimum QDI early output RCA (i.e., O8), ii) 14.2% and 13.6% reductions in CT and PCTP respectively compared to an optimum relative-timed RCA (i.e., O9), iii) 31.7% and 34.8% reductions in CT and PCTP respectively compared to an optimum uniform input-partitioned QDI early output CSLA (i.e., O10), iv) 47.1% and 46.8% reductions in CT and PCTP respectively compared to an optimum QDI early output CCLA (i.e., O16), v) 44.7% and 44.4% reductions in CT and PCTP respectively compared to an optimum QDI early output BCLA (i.e., O19), vi) 13% and 12.4% reductions in CT and PCTP respectively compared to an optimum QDI early output BCLARC (i.e., O20), and vii) 11.1% and 10.5% reductions in CT and PCTP respectively compared to an optimum QDI early output hybrid BCLARC-RCA (i.e., O21).

Further work would be to investigate the usefulness of the proposed QDI BCLARC in realizing other computer arithmetic operations of practical significance by considering accurate and approximate computations.

## Acknowledgment

This research was funded by the Academic Research Fund Tier-2 research award of the Ministry of Education (MOE), Singapore under Grant MOE2017-T2-1-002.